\documentclass[useAMS,usenatbib]{mn2e}
\usepackage{multirow}
\usepackage{epsfig}
\usepackage{subfigure}

\makeatletter
\let\jnl@style=\rm
\def\ref@jnl#1{{\jnl@style#1}}

\def\aj{\ref@jnl{AJ}}                   
\def\araa{\ref@jnl{ARA\&A}}             
\def\apj{\ref@jnl{ApJ}}                 
\def\apjl{\ref@jnl{ApJ}}                
\def\apjs{\ref@jnl{ApJS}}               
\def\ao{\ref@jnl{Appl.~Opt.}}           
\def\apss{\ref@jnl{Ap\&SS}}             
\def\aap{\ref@jnl{A\&A}}                
\def\aapr{\ref@jnl{A\&A~Rev.}}          
\def\aaps{\ref@jnl{A\&AS}}              
\def\azh{\ref@jnl{AZh}}                 
\def\baas{\ref@jnl{BAAS}}               
\def\jrasc{\ref@jnl{JRASC}}             
\def\memras{\ref@jnl{MmRAS}}            
\def\mnras{\ref@jnl{MNRAS}}             
\def\pra{\ref@jnl{Phys.~Rev.~A}}        
\def\prb{\ref@jnl{Phys.~Rev.~B}}        
\def\prc{\ref@jnl{Phys.~Rev.~C}}        
\def\prd{\ref@jnl{Phys.~Rev.~D}}        
\def\pre{\ref@jnl{Phys.~Rev.~E}}        
\def\prl{\ref@jnl{Phys.~Rev.~Lett.}}    
\def\pasp{\ref@jnl{PASP}}               
\def\pasj{\ref@jnl{PASJ}}               
\def\qjras{\ref@jnl{QJRAS}}             
\def\skytel{\ref@jnl{S\&T}}             
\def\solphys{\ref@jnl{Sol.~Phys.}}      
\def\sovast{\ref@jnl{Soviet~Ast.}}      
\def\ssr{\ref@jnl{Space~Sci.~Rev.}}     
\def\zap{\ref@jnl{ZAp}}                 
\def\nat{\ref@jnl{Nature}}              
\def\iaucirc{\ref@jnl{IAU~Circ.}}       
\def\aplett{\ref@jnl{Astrophys.~Lett.}} 
\def\apspr{\ref@jnl{Astrophys.~Space~Phys.~Res.}}
\def\bain{\ref@jnl{Bull.~Astron.~Inst.~Netherlands}}
\def\fcp{\ref@jnl{Fund.~Cosmic~Phys.}}  
\def\gca{\ref@jnl{Geochim.~Cosmochim.~Acta}}   
\def\grl{\ref@jnl{Geophys.~Res.~Lett.}} 
\def\jcp{\ref@jnl{J.~Chem.~Phys.}}      
\def\jgr{\ref@jnl{J.~Geophys.~Res.}}    
\def\jqsrt{\ref@jnl{J.~Quant.~Spec.~Radiat.~Transf.}}
\def\memsai{\ref@jnl{Mem.~Soc.~Astron.~Italiana}}
\def\nphysa{\ref@jnl{Nucl.~Phys.~A}}   
\def\physrep{\ref@jnl{Phys.~Rep.}}   
\def\physscr{\ref@jnl{Phys.~Scr}}   
\def\planss{\ref@jnl{Planet.~Space~Sci.}}   
\def\procspie{\ref@jnl{Proc.~SPIE}}   

\def \nh {N${\rm _H}$}

\def \arcmin {\hbox{$^\prime$}}
\def \arcsec {\hbox{$^{\prime\prime}$}}

\def\spose#1{\hbox to 0pt{#1\hss}}
\def\ltsim{$\mathrel{\spose{\lower 3pt\hbox{$\sim$}}
        \raise 2.0pt\hbox{$<$}}$\thinspace}
\def\gtsim{$\mathrel{\spose{\lower 3pt\hbox{$\sim$}}
        \raise 2.0pt\hbox{$>$}}$\thinspace}
\def \msun {${\rm M_\odot}$}

\newcommand\solar{\hbox{{$Z_{\odot}$}}}

\def \nh {$N_{\rm H}$}

\newcommand{\source}{\mbox{Abell\,3783}}

\newcommand{\apec}{APEC}

\newcommand{\chandra }{{\em Chandra}}
\newcommand{\xspec }{{\em Xspec}}

\newcommand{\fxunits}{\mbox{ergs cm$^{-2}$ s$^{-1}$}}
\newcommand{\lxunits}{\mbox{ergs s$^{-1}$}}
\newcommand{\xmm }{XMM-{\it Newton}}

\newcommand{\rosat }{{\em ROSAT}}

\newcommand{\lx }{${\rm L_X}$}
\newcommand{\tx }{${\rm T_X}$}

\newcommand{\lsun }{${\rm L_\odot}$}

\newcommand{\ned}{{\em{NED}}}

\makeatother

\title[Unabsorbed Seyfert 2 galaxies: the case of ``naked'' AGN]{Unabsorbed Seyfert 2 galaxies: the
case of ``naked'' AGN\thanks{Based on observations obtained with XMM-\textit{Newton} and at ESO NTT telescope
(La Silla, Chile, program: 278.B-5021).}}
\author[Francesca Panessa et al.]{F. Panessa$^1$\thanks{E-mail: francesca.panessa@iasf-roma.inaf.it},
F.J. Carrera$^2$, S. Bianchi$^3$, A. Corral$^4$, F. Gastaldello$^5\,^6\,^7$,
\newauthor 
X. Barcons$^2$, L. Bassani$^8$, G. Matt$^3$, L. Monaco$^9$,\\
$^1$Istituto di Astrofisica Spaziale e Fisica Cosmica (IASF-INAF), via del Fosso del Cavaliere 100, 00133 Roma, Italy\\
$^2$Instituto de F\'\i sica de Cantabria (CSIC-UC), 39005 Santander, Spain\\
$^3$Dipartimento di Fisica, Universit\`a degli Studi Roma Tre, via della Vasca Navale 84, 00146 Roma, Italy\\
$^4$Osservatorio Astronomico di Brera (OAB-INAF), via Brera 28, 20121 Milano, Italy\\
$^5$IASF-Milano, INAF, via Bassini 15, Milano 20133, Italy\\
$^6$Department of Physics and Astronomy, University of California at Irvine, 4129, Frederick Reines Hall, Irvine, CA 92697-4575\\
$^7$Occhialini Fellow\\
$^8$Istituto di Astrofisica Spaziale e Fisica Cosmica (IASF-INAF), Via P. Gobetti 101, 40129 Bologna, Italy\\
$^9$Universidad de Concepci\'on, Casilla 160-C, Concepci\'on, Chile\\ 
}
\begin{document}

\date{}

\pagerange{\pageref{firstpage}--\pageref{lastpage}} \pubyear{2002}

\maketitle

\label{firstpage}

\begin{abstract}
Hawkins (2004) reported on a class of ``naked" AGN characterized 
by strong amplitude optical brightness variability and the complete absence 
of broad emission lines in the optical spectrum. 
The variability suggests that the nucleus is seen directly, however
the absence of broad lines contradicts the simple formulation of Unified Models for AGN.
We present the results of quasi-simultaneous spectroscopic observations with XMM-\textit{Newton} and
NTT ({\it La Silla}) of two ``naked" AGN. We confirm the ``naked" nature of Q\,2131-427
for which no broad emission line components have been detected in the optical spectrum
and its X-ray spectrum shows no signs of intrinsic absorption. 
The optical and X-ray mismatch in this source cannot be ascribed to 
a high nuclear dust-to-gas ratio and a Compton Thick nature is ruled out
on the basis of the high F$_{X}$/F$_{[\textsc{oiii}]}$ ratio.
The Broad Line Region (BLR) may be completely absent in this source, possibly as a consequence of its low Eddington ratio.
On the other hand, the optical spectrum of Q\,2130-431 shows H$_{\alpha}$ and H$_{\beta}$
broad emission line components, revealing the presence of a BLR. A mild X-ray absorption
is expected in intermediate type 1.8 Seyfert galaxies like Q\,2130-431, however we put a 
very low upper limit on the column density ($<$ 2 $\times$ $10^{20}$ cm$^{-2}$), also the low
Balmer decrement suggests that the BLR itself does not suffer from reddening. We propose that
in this object the BLR is intrinsically weak, making a case of ``true" intermediate Seyfert galaxy.
We also report on the X-ray detection of the Abell 3783 galaxy cluster in the 
XMM-\textit{Newton} field-of-view of the Q\,2131-427 observation.
\end{abstract}


\section{Introduction}

The Unified Model for Seyfert galaxies predicts that 
the differences observed between type 1 and type 2 Seyfert
galaxies are primarily due to orientation effects (Antonucci 1993).
Optical narrow emission lines present in both type 1 and 
type 2 Seyfert's spectra are produced in the Narrow Line Region (NLR)
at $> 100$ pc scale from the nucleus. Optical broad emission lines 
originate in the Broad Line Region (BLR) at sub-pc scale. The latter
are observed only in type 1 Seyfert's spectra since, in type 2 Seyferts,
they are obscured by a molecular torus.
Much evidence has been found in favor of this picture,
such as the larger amount of absorbing material
measured from X-ray observations in Seyfert 2s with respect
to Seyfert 1s (Risaliti et al. 1999, Awaki et al. 1991).

However, in the last few years the number of
cases in which observations do not match with Unified Models
is increasing both in the local and in the distant universe.
Type 1 Active Galactic Nuclei (AGN) with significant absorption have been found 
(Cappi et al. 2006, Mateos et al. 2005, Fiore et al. 2001)
as well as type 2 AGN without X-ray absorption
(Brightman \& Nandra 2008, Bianchi et al. 2008, Wolter at al. 2005, Corral et al. 2005, Caccianiga et al. 2004, Barcons, 
Carrera \& Ceballos 2003, Panessa \& Bassani 2002,
Pappa et al. 2001). 

\begin{table*}
\label{sam}
\begin{center}
\caption{\bf The ``naked" AGN sample and XMM-\textit{Newton} Observation Log}
\begin{tabular}{lccccccccccccc}
\hline
\hline
\multicolumn{1}{c}{Name} &
\multicolumn{1}{c}{RA} &
\multicolumn{1}{c}{DEC} &
\multicolumn{1}{c}{z} &
\multicolumn{1}{c}{[O\textsc{iii}]/H$_{\beta}$} &
\multicolumn{1}{c}{$\delta$B} &
\multicolumn{1}{c}{Obs. Date} &
\multicolumn{1}{c}{Exposure (s)} &
\multicolumn{1}{c}{Filters}  \\
\multicolumn{1}{c}{} &
\multicolumn{1}{c}{} &
\multicolumn{1}{c}{} &
\multicolumn{1}{c}{} &
\multicolumn{1}{c}{} &
\multicolumn{1}{c}{} &
\multicolumn{1}{c}{M1/M2/PN} &
\multicolumn{1}{c}{M1/M2/PN} &
\multicolumn{1}{c}{M1/M2/PN}  \\
\multicolumn{1}{c}{(1)} &
\multicolumn{1}{c}{(2)} &
\multicolumn{1}{c}{(3)} &
\multicolumn{1}{c}{(4)} &
\multicolumn{1}{c}{(5)} &
\multicolumn{1}{c}{(6)} &
\multicolumn{1}{c}{(7)} &
\multicolumn{1}{c}{(8)} &
\multicolumn{1}{c}{(9)}  \\
\hline
\hline
Q\,2130-431 & 21 33 15.62 & -42 54 24 &	0.266 & 16.36 & 1.20 &  2006-11-13 & 29997/29774/24431 & M/M/M \\  
Q\,2131-427 & 21 34 26.49 & -42 29 56 &	0.365 &  7.43 & 1.18 &  2006-11-15 & 23698/23429/16140 & M/M/M \\   
\hline
\hline
\end{tabular}
\end{center}
Notes: Col (1): Galaxy name; col (2)-(3) Optical position in epoch J2000;
col. (4) Redshift; col. (5) [O\textsc{iii}]/H$_{\beta}$ ratio; col. (6): Amplitude
of the $B_{J}$ band measured over the period from 1974 to 2002; col. (7): Observation date; 
col. (8): MOS1/MOS2/PN observation exposures; col. (9): MOS1/MOS2/PN filters, M=medium.
Columns (1), (4)-(6) are from Hawkins (2004).
\end{table*}

Several explanations have been proposed to reconcile 
the Unified Model paradigm with these pieces of evidence. 
For instance, the broad emission lines could fade away
in response to a decrease of the continuum emission (Guainazzi et al. 2005, Matt et al. 2003).
Alternatively, the BLR is covered by clumpy and variable obscuring material,
as in NGC\,4388 where a variation of a factor of 100 in column density
has been observed from X-ray observations (Elvis et al. 2004)
and in NGC\,1365 where a spectral change from Compton-thin to Compton-thick and back to Compton-thin 
has happened in four days (Risaliti et al. 2007). In the above mentioned cases,
the misleading X-ray and optical behavior is basically due to non-simultaneity
of the observations.
Finally, the BLR could be weak or absent and
its formation linked to the accretion physics (Wang \& Zhang 2007, Elitzur \& Shlosman 2006, Nicastro et al. 2003,
Nicastro 2000, Williams et al. 1999).
Simultaneous optical and X-rays observations have confirmed
that NGC\,3147 is BLR-free and without X-ray absorption (Bianchi et al. 2008).  
The strategy of observing the source simultaneously in the two bands
has been proved to be fundamental in order 
to avoid mismatches due to spectral or flux variability of the source at X-rays and optical wavebands.

Hawkins (2004) presented the results from
a long term monitoring of about 800 quasars. Among
them a new class of AGN is reported, i.e. the ``naked" AGN, where the absence of broad
emission lines is accompanied by strong optical variability, suggesting
that the nucleus is seen directly. Subsequently \textit{Chandra} snapshot observations
of three ``naked" AGN (Q\,2130-431, Q\,2131-427 and Q\,2122-444) by Gliozzi et al. (2007) confirmed this hypothesis given
the absence of significant absorption in the X-ray spectra, though of low statistics.
In this work we present quasi-simultaneous X-ray and optical spectroscopic observations
with XMM-{\it Newton} and EMMI/NTT ({\it La Silla Observatory}, LSO, ESO) of two sources (Q\,2130-431 and Q\,2131-427)
out of the six ``naked" AGN sample defined in Hawkins (2004) and discuss their nature with respect to Unified
Models and recent developments.
We also report on the serendipitous detection of the cluster \source\ in Appendix A.
In the following, we adopt $H_0=70$
km s$^{-1}$ Mpc$^{-1}$, $\Omega_\Lambda=0.73$ and $\Omega_m=0.27$ (Spergel et al. 2003).

\section{The sample of ``naked" AGN}

Hawkins (2004) carried out a yearly
photometrical large-scale monitoring programme for AGN over the last 25 years.
The survey was based on a long series of photographic plates from the UK 1.2m Schmidt telescope.
Candidate AGN were selected from a catalogue of 200,000 objects to perform follow-up spectroscopic study.
Optical spectra (obtained in July 2002) pinpoint a class of objects 
(6 in that sample) which show narrow and weak H$_{\beta}$
emission line, large [O\textsc{iii}]$\lambda$5007/H$_{\beta}$ ratios
typical of type 2 Seyferts (i.e., [O\textsc{iii}]$\lambda$5007/H$_{\beta}$ $>$ 3, Shuder \& Osterbrock 1981), 
and no sign of broad emission lines\footnote{Note, however, that the broadness of other emission lines as, 
for instance, H$_{\alpha}$, could not be determined since this line was redshifted out of the optical passband.}.
For this class of objects, the difference between the maximum and minimum light in magnitudes in the B$_{J}$ passband
over a period of 25 years, $\delta$B, reveals large amplitude variations normally only found in type 1 objects.
Their brightness varied at least by a factor of 3 on a timescale
of 5-10 years and also on shorter time scales (see light curves in Hawkins 2004).
Two among the brightest sources of the six ``naked" AGN sample have been observed in this work (Q\,2130-431 and Q\,2131-427).
In particular, Hawkins (2004) reported also on a previous optical
observation of Q\,2131-427, in 1991 with EFOSC on the 3.6m  at ESO, when the source was
0.7 magnitudes brighter than in 2002. Its optical spectrum showed no sign of broad emission
lines and weak H$_{\beta}$, basically consistent with the spectrum observed afterward in 2002,
apart from some additional flux at the blue end of the continuum. 
In Table~\ref{sam} we report the two observed ``naked" AGN, where
redshift, [O\textsc{iii}]$\lambda$5007/H$_{\beta}$ and $\delta$B are taken from Hawkins (2004).

\section{Data reduction and analysis}

\subsection{Optical data}

\begin{table*}
\caption{\bf Measured line parameters for the optical spectrum of Q\,2130-431 and Q\,2131-427.}
\small{
\begin{center}
\begin{tabular}{lrlrl}
\hline
\hline
\multicolumn{1}{c}{} &
\multicolumn{2}{c}{Q\,2130-431} &
\multicolumn{2}{c}{Q\,2131-427} \\
\hline
\multicolumn{1}{c}{Line} &
\multicolumn{1}{c}{Flux ($10^{-15}$ cgs) } &
\multicolumn{1}{c}{FWHM (km s$^{-1}$)} &
\multicolumn{1}{c}{Flux ($10^{-15}$ cgs)} &
\multicolumn{1}{c}{FWHM (km s$^{-1}$)}  \\

\hline
\hline
H$_\beta$ (narrow)		            & 0.11 & 320$^{+50}_{-30}$ 	  (426)$\dagger$	& 0.13 & 590$^{+40}_{-60}$ (830)$\dagger$	 \\
H$_\beta$ (broad) 			    & 1.17 & 9680$^{+190}_{-150}$ 			& $<$0.06 & 4000$^{*}$     	\\
$\mathrm{[{O\,\textsc{iii}}]}\,\lambda4959$ & 0.40 & 370$^{+30}_{-20}$ 	  			& 0.40 & 970$^{+40}_{-50}$ 	\\
$\mathrm{[{O\,\textsc{iii}}]}\,\lambda5007$ & 1.19 & 370 	(410)$\dagger$ 			& 1.21 & 970 (960)$\dagger$ 	 \\
H$_\alpha$ (narrow) 			    & 0.39 & 320   					& 0.54 & 590     \\
H$_\alpha$ (broad)                          & 4.01 & 9680 					& $<$0.25 & 4000$^{*}$\\   
$\mathrm{[{N\,\textsc{ii}}]}\,\lambda6548$  & 0.22 & 440$^{+20}_{-30}$    			& 0.24 & 890$^{+60}_{-70}$     \\
$\mathrm{[{N\,\textsc{ii}}]}\,\lambda6583$  & 0.66 & 440		   			& 0.72 & 890    \\
$\mathrm{[{S\,\textsc{ii}}]}\,\lambda6716$  & 0.10 & 320$^{+50}_{-30}$    			& - & - 		       \\
$\mathrm{[{S\,\textsc{ii}}]}\,\lambda6731$  & 0.10 & 320		   			& - & - 		       \\
\hline
\end{tabular}
\end{center}
Notes: Line widths for [O\textsc{iii}], [N\textsc{ii}], [S\textsc{ii}] doublets and Balmer line components (narrow and broad respectively)
were fitted by tying down the velocity widths to the same value. Parameters with $^*$ were kept fixed
to the reported value during the fitting procedure. $\dagger$ FWHM in km s$^{-1}$ obtained from Hawkins (2004)
after correction for the spectral resolution of 520 km s$^{-1}$ (see Hawkins (2004), sec 2.2).} 
\label{table=opt_spec}
\end{table*}

Q\,2130-431 and Q\,2131-427 were observed at the  NTT telescope in La Silla (Chile) in
director discretionary time (ID: 278.B-5021) to obtain a quasi-simultaneous observation 
with the XMM-{\it Newton} ones. Observations for both sources have been performed on 2006, December 20th with 
the ESO Multi-Mode Instrument (EMMI) with grating \#4 and a 1-arcsec
slit width. The condition of the night was good (i.e., clear to photometric).
The exposure for each spectrum was of 1200s (air mass 1.70 for Q\,2130-431 and 1.92 for Q\,2131-427).  
The wavelength resolution measured from unblended arc lines was 3 \AA, in the
range 6000-9000 \AA. A standard reduction process was applied using midas and iraf tasks. The raw data
were bias subtracted and corrected for pixel-to-pixel variations (flat-field). Object spectra were extracted
and sky subtracted.
Wavelength calibrations were carried out by comparison with exposures of He and Ar lamps, with an accuracy 
$<$ 1 \AA. Relative flux calibration was carried out by observations of the spectrophotometric standard
star SA95\_42 (Oke 1990). We estimate an error on the flux calibration $<$ 10\% from the standard
adjustment during the calibration procedure. Statistical errors were propagated throughout the data reduction process.

The spectral fitting was carried out using the QDP fitting routines via $\chi^{2}$ minimisation. 
The spectrum has been fitted with a linear component in order to reproduce the continuum which is 
always almost flat.
The line intensities and widths were measured by fitting Gaussian profiles and the Full Width Half Maximum (FWHM)
calculated taking into account the spectrograph resolution. The NLR is stratified as it is the BLR, 
but emission lines coming from the same region (like the doublets and the Balmer lines) should display similar widths. 
For this reason, velocity widths for narrow lines were tied together to the same value
for each doublet as well as for the narrow components (and broad, if any) of the Balmer lines. The results from the spectral fitting
of the two objects are shown in Table~\ref{table=opt_spec} and the optical fitted spectra are shown in 
Figures~\ref{opt1} and~\ref{opt2}. 

In the case of Q\,2130-431 a broad component with a FWHM $\sim$ 9700 km s$^{-1}$ is clearly
detected in the H$_{\alpha}$ and H$_{\beta}$ emission lines. According to the Whittle (1992) 
classification criteria for intermediate type Seyfert galaxies (based on the F$_{[\textsc{oiii}]}$/F$_{H\beta (broad)}$ ratio),
Q\,2130-431 should be classified as Seyfert 1.5. However, F$_{[\textsc{oiii}]}$/F$_{H\beta (broad)}$ = 1.0
is close to the boundary between the Seyfert 1.5 and Seyfert 1.8 classification; since
the Seyfert 1.5 classification is mainly used for sources with highly variable spectra, 
we adopt a Seyfert 1.8 classification for Q\,2130-431.

Conversely, no broad components (H$_{\alpha}$ nor H$_{\beta}$) are significantly detected in the spectra of Q\,2131-427. 
We compute an upper limit on the H$_{\alpha}$ and H$_{\beta}$ broad emission line components
by fixing the line widths to 4000 km s$^{-1}$, a conservative value for BLR widths (Ho et al. 1997). 
We are confident that in this source the lack of detection of broad line components is not due to the host galaxy
contamination. Indeed, our spectra are quite flat and do not show any significant contribution from the host galaxy. 
Since the data are taken in long-slit mode, we are able to extract only the most central part of the source, 
thus avoiding as much as possible the host galaxy contribution. 

We calculate the fractional contribution of the H$_{\alpha}$ broad component
to the total (broad$+$narrow) H$_{\alpha}$ emission by using the estimated upper limit. 
This corresponds to at most 30 per cent, a low value if compared to the average value 
in the Ho et al. (1997) Seyfert sample which is around 65 per cent. Note that the latter is a 
conservative estimate since the average FWHM for broad line components in Ho et al. (1997) is around 2700 km s$^{-1}$ while
our upper limit has been calculated for a FWHM of 4000 km s$^{-1}$.

In the classical [O\textsc{iii}]/H$_{\beta}$ vs. [NII]/H$_{\alpha}$ diagnostic diagram plot 
used to classify narrow emission line galaxies (Kewley et al. 2006), both sources are placed in the Seyfert region, well above the LINER-Starburst
limits. In conclusion, the optical classification for Q\,2130-431 and Q\,2131-427 is Seyfert 1.8 and Seyfert 2,
respectively.  

It is interesting to compare our spectral results with those obtained from Hawkins (2004) and here reported within parenthesis
in Table~\ref{table=opt_spec}. The [O\textsc{iii}] and H$_{\beta}$ FHWM measured for Q\,2130-431
are consistent with the ones reported in Hawkins (2004), after spectral resolution correction,
while our measured [O\textsc{iii}]/H$_{\beta}$ ratio of $\sim$ 11 is smaller
compared to $\sim$ 16 (see Table~\ref{sam}). 
In the optical spectrum of Q\,2131-427 the major difference results in the H$_{\beta}$ FWHM significantly smaller 
than the one measured by Hawkins (2004) ($\sim$ 830 km s$^{-1}$). On the other hand, the [O\textsc{iii}]/H$_{\beta}$ ratio $\sim$ 9
and the [O\textsc{iii}] width are consistent within errors. We note that, this comparison exercise should be taken
with caution, since the quality of the spectroscopy observations in Hawkins (2004), performed with 2dF multi-fibre spectrograph, 
cannot provide accurate error estimates for the FWHM and could also be problematic when calculating line ratios, given
the unreliability of sky subtraction and flux calibration.   

\begin{figure*}
\epsfig{figure=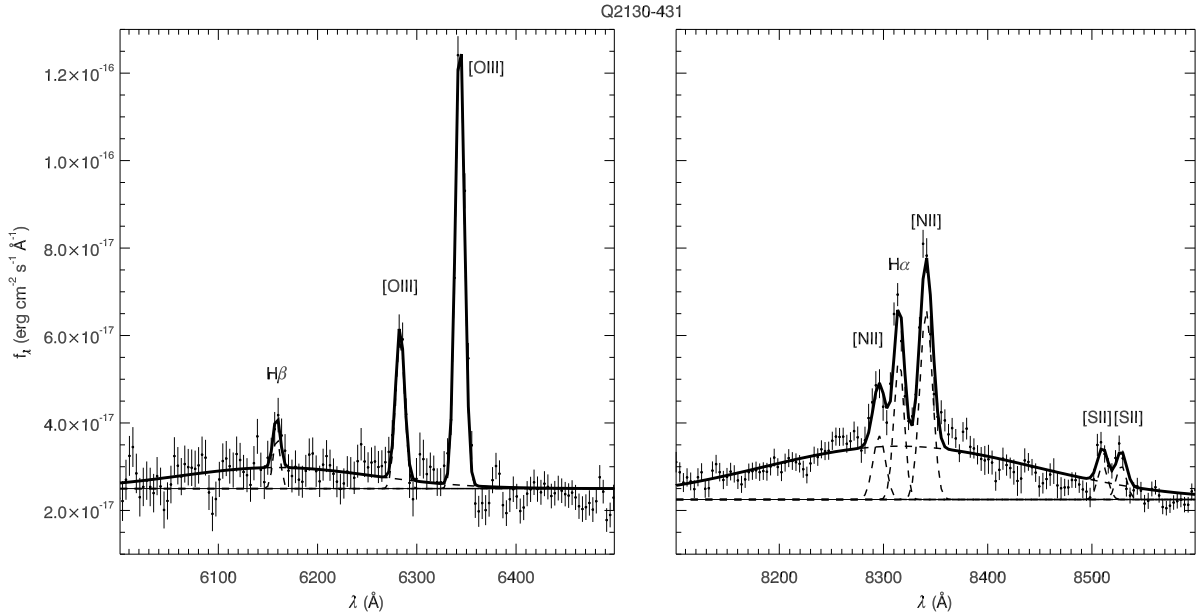,height=8.2cm,angle=0}
\caption{\label{opt1} Q\,2130-431 ESO EMMI optical spectrum. Best-fitting models
of the H$_{\beta}$ (left panel) and H$_{\alpha}$ (right panel) spectral regions.}
\label{opt1}
\end{figure*}

\begin{figure*}
\epsfig{figure=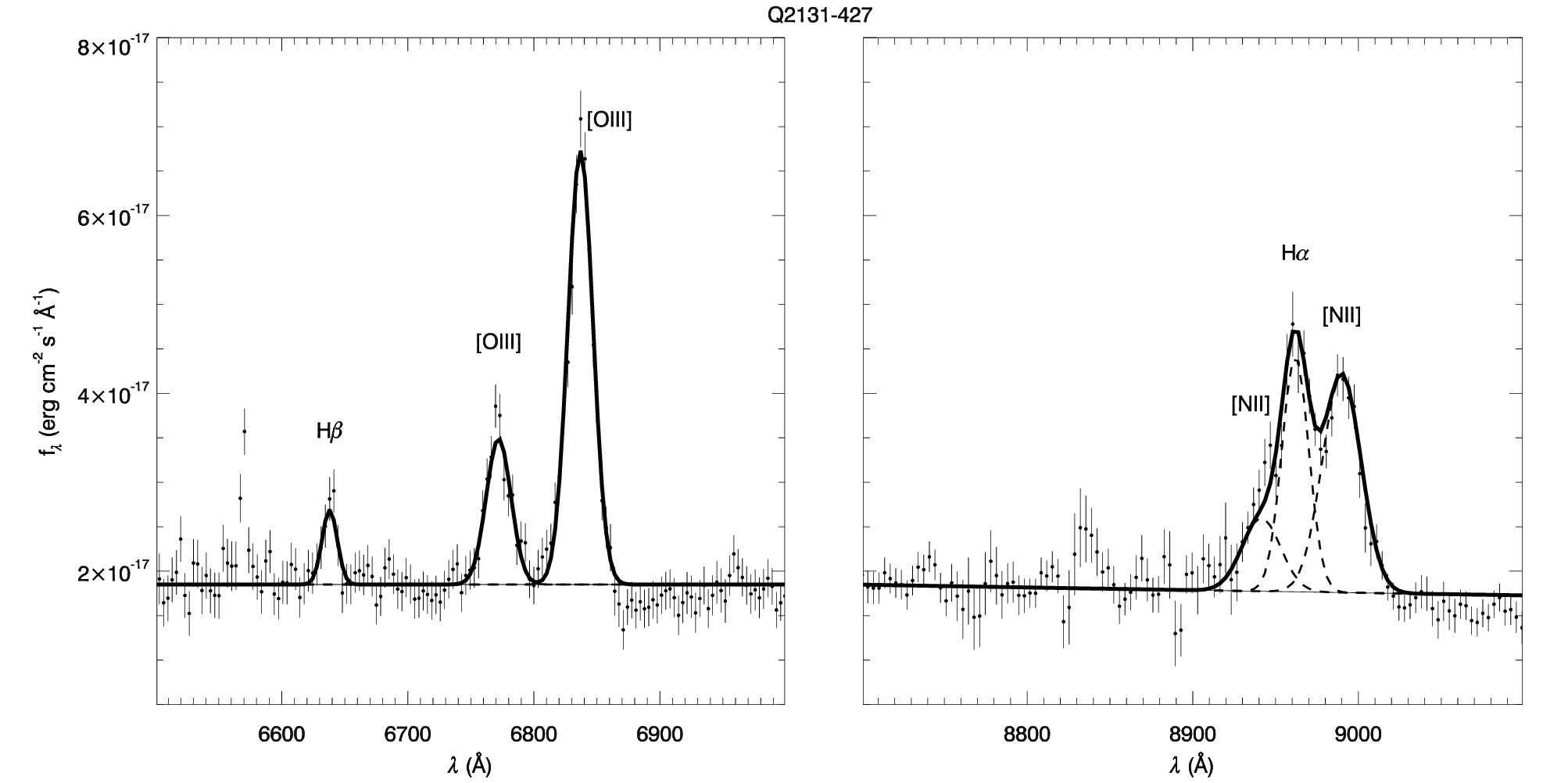,height=8cm,angle=0}
\caption{\label{opt2} Q\,2131-427 ESO EMMI optical spectrum. Best-fitting models
of the H$_{\beta}$ (left panel) and H$_{\alpha}$ (right panel) spectral regions.}
\end{figure*}

\begin{table*}
\caption{\bf XMM-\textit{Newton} spectral parameters, black hole mass and Eddington ratio.}
\small{
\begin{center}
\begin{tabular}{lcccccccccc}
\hline
\hline
\multicolumn{1}{l}{Name} &
\multicolumn{1}{c}{$N_{\rm H, Gal}$} &
\multicolumn{1}{c}{$N_{\rm H, int}$} &
\multicolumn{1}{c}{$\Gamma$} &
\multicolumn{1}{c}{$\chi^{2}/dof $}  &
\multicolumn{1}{c}{$F_{\rm 0.5-2}$}  &
\multicolumn{1}{c}{$F_{\rm 2-10}$} &
\multicolumn{1}{c}{$L_{\rm  0.5-2}$}  &
\multicolumn{1}{c}{$L_{\rm  2-10}$}  &
\multicolumn{1}{c}{M$_\mathrm{BH}$} &
\multicolumn{1}{c}{L$_\mathrm{Bol}$/L$_\mathrm{Edd}$} \\
\multicolumn{1}{c}{} &
\multicolumn{1}{c}{} &
\multicolumn{1}{c}{} &
\multicolumn{1}{c}{} &
\multicolumn{1}{c}{}  &
\multicolumn{1}{c}{(cgs)}  &
\multicolumn{1}{c}{(cgs)} &
\multicolumn{1}{c}{(cgs)}  &
\multicolumn{1}{c}{(cgs)} &
\multicolumn{1}{c}{$M_\odot$} &
\multicolumn{1}{c}{}  \\
\multicolumn{1}{c}{(1)} &
\multicolumn{1}{c}{(2)} &
\multicolumn{1}{c}{(3)} &
\multicolumn{1}{c}{(4)} &
\multicolumn{1}{c}{(5)} &
\multicolumn{1}{c}{(6)} &
\multicolumn{1}{c}{(7)} &
\multicolumn{1}{c}{(8)} &
\multicolumn{1}{c}{(9)} &
\multicolumn{1}{c}{(10)} &
\multicolumn{1}{c}{(11)}  \\
\hline
\hline
Q\,2130-431 & 2.4 & $<$ 2.0 & 1.77$^{+0.04}_{-0.04}$ & 149.8/144 & 7.9$\times$10$^{-14}$ & 1.5$\times$10$^{-13}$ & 2.0$\times$10$^{43}$ & 3.8$\times$10$^{43}$ &
1.6$\times$10$^{7}$ & 0.37 \\
Q\,2131-427 & 2.2 & $<$ 9.0 & 2.00$^{+0.23}_{-0.21}$ & 27.3/23   & 1.8$\times$10$^{-14}$ & 2.2$\times$10$^{-14}$ & 1.0$\times$10$^{43}$ & 1.2$\times$10$^{43}$ &
7.7$\times$10$^{8}$ & 2.4$\times$10$^{-3}$ \\
\hline
\end{tabular}
\end{center}
Notes: Col. (1): Name of the source; col. (2): Galactic column density in units of $10^{20}$ (cm$^{-2})$; 
col. (3): Intrinsic column density in units of $10^{20}$ (cm$^{-2})$; col. (4) Photon index; col. (5) Chi-squared and degrees of freedom;
col. (6): Intrinsic flux in the 0.5--2 keV band; 
col. (7): Intrinsic flux in the 2--10 keV band; col. (8): Intrinsic luminosity in the 0.5--2 keV band; 
col. (9): Intrinsic luminosity in the 2--10 keV band; col. (10): black hole mass; col. (11): Ratio of bolometric to Eddington luminosity. 
} 
\label{table=obs_spec}
\end{table*}

\begin{figure*}
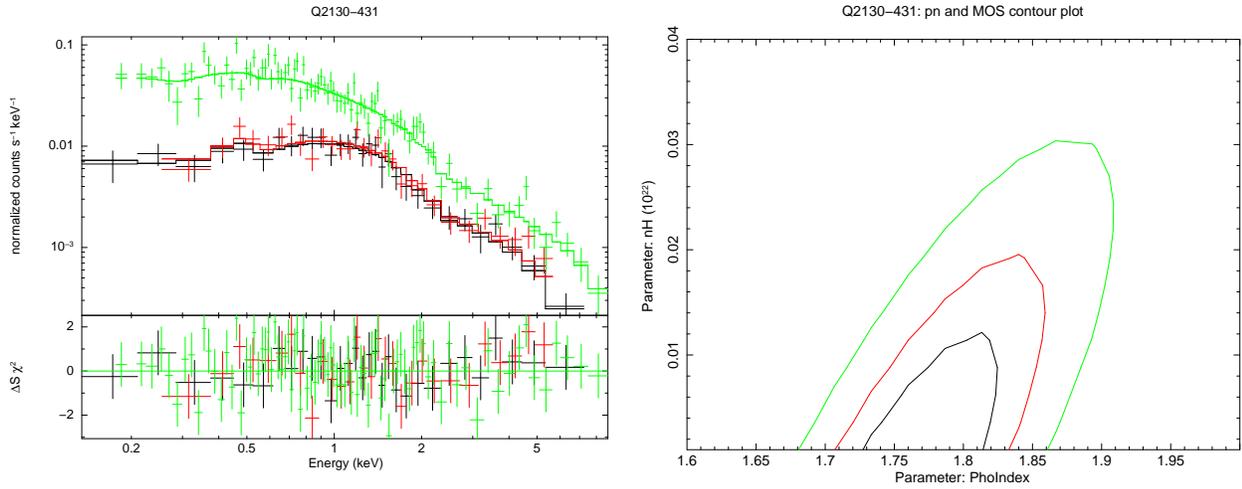

\epsfig{figure=new_Q2130-431.ps,height=8cm,angle=-90}
\hspace{0.2cm}
\epsfig{figure=new_wawapo_cont.ps,height=8cm,angle=-90}
\caption{\label{Q2130} XMM-\textit{Newton} observation of Q\,2130-431: \textit{Left panel}: The EPIC pn, MOS1 and MOS2 spectra, 
best fit models and $\Delta\chi^2$ deviations. \textit{Right panel}: Column density versus photon index contour 
plots at 68, 90 and 99\% confidence level for two interesting parameters.}
\end{figure*}

\begin{figure*}
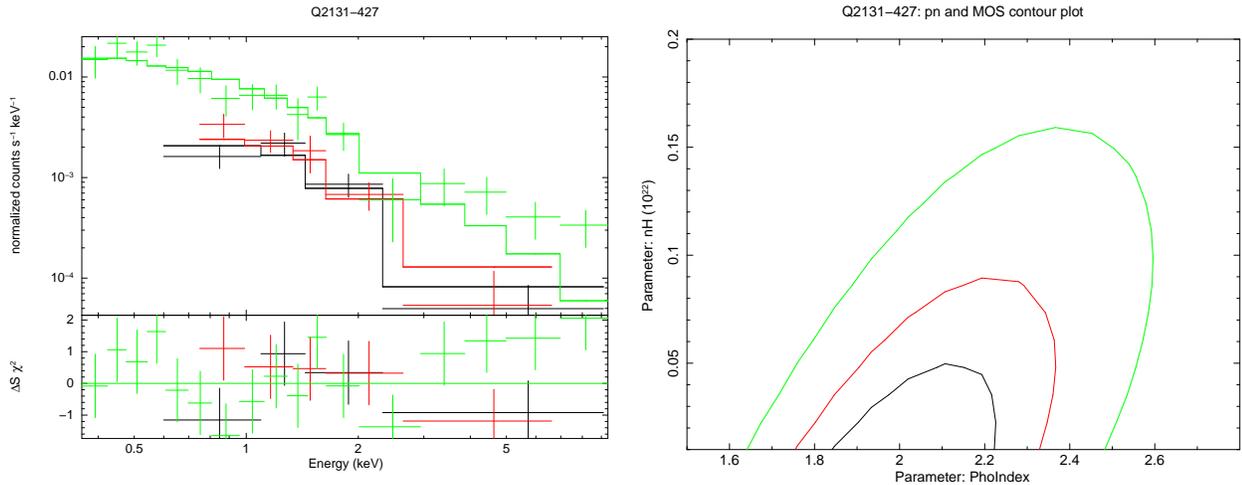

\epsfig{figure=Q2131_new.ps,height=8cm,angle=-90}
\hspace{0.2cm}
\epsfig{figure=Q2131_new_cont.ps,height=8cm,angle=-90}
\caption{\label{Q2131} XMM-\textit{Newton} observation of Q\,2131-427: \textit{Left panel}: The EPIC pn, MOS1 and MOS2 spectra, 
best fit models and $\Delta\chi^2$ deviations. \textit{Right panel}: Column density versus photon index contour 
plots at 68, 90 and 99\% confidence level for two interesting parameters.}
\end{figure*}

\subsection{XMM-\textit{Newton} Observations}

Q\,2130-431 and Q\,2131-427 were observed by XMM-\textit{Newton} on November 2006
(\textsc{obsid: 0402460201}). Data have been processed starting from the observation data
fits files
with \textsc{SAS} 7.0.0. X-ray events 
corresponding to patterns 0-12 and 0-4 were selected from
the EPIC-MOS and EPIC-pn cameras, respectively. We used the most updated calibration
files available at the time of the reduction for each source data.
Source light curves and spectra were extracted from circular regions of
25$\arcsec$ centered on the source, while background products were obtained from 
off-set regions close to the source. Exposure times have been
filtered out for periods of high background and the ``good time interval"
exposures are reported in Table~\ref{sam} as well as
the observation date and the instruments filters. 
Spectra were binned according to the counts collected in each source.
The ancillary and detector response matrices were generated using
the XMM-\textit{Newton} SAS \textsc{arfgen} and \textsc{rmfgen} tasks.
The X-ray light-curves have been examined and no significant variability has been found.
 
The pn, MOS1 and MOS2 spectra were analyzed using XSPEC v.12.4.0. 
Galactic absorption is implicitly included in all spectral models.
Abundances are those of Anders \& Grevesse (1989).
The errors, lower and upper limits quoted correspond to 90\% confidence range for one
interesting parameter (i.e., $\Delta\chi^2 = 2.71$; Avni 1976), unless otherwise stated.

The X-ray spectrum of Q\,2130-431 is well represented by
a single power-law with photon spectral index $\Gamma=1.77\pm0.04$ ($\chi^{2}/\nu$ = 150/144).
No extra absorption apart from the Galactic one is needed 
($N_{\rm H, int}$ $<$ 2 $\times$ $10^{20}$ cm$^{-2}$).
Q\,2131-427 shows a very weak spectrum, modeled by a
single power-law ($\Gamma$ $\sim$ 2) and an upper limit (at 90\% confidence level) on the
intrinsic absorption of $N_{\rm H, int}$ $<$ 9 $\times$ $10^{20}$ cm$^{-2}$.
We fixed a narrow Gaussian emission line at the rest energy of the FeK$_{\alpha}$ line, 6.4 keV, and we measured
an upper limit on the equivalent width of the line of $\sim$ 2 keV. However,
this large value is probably due to the significant residuals visible
above 5 keV in the pn spectrum; actually the very low photon statistics around 6-7 keV
and the background dominance in this range prevent us from drawing any strong conclusions based on this limit.
In Figures~\ref{Q2130} and ~\ref{Q2131} best fit spectra are shown on the left panels,
while column density versus photon index contour plots are on the right panels.
In Table~\ref{table=obs_spec} we show the best--fit parameters 
together with the model fluxes and luminosities in the 0.5-2 keV and 2-10 keV bands. 
We also report the black hole mass derived from the $M_{BH}$ $-$ $\sigma_{\star}$ relation (Tremaine
et al. 2002), where $\sigma_{\star}$ is the stellar velocity dispersion
obtained from the FWHM([O\textsc{iii}]) as shown in Greene \& Ho (2005). The Eddington
ratio has been calculated assuming that $L_{Bol}/L_{X}$ $=$ 20 (Elvis et al. 1994, Vasudevan \& Fabian 2009).

\section{Discussion}

The ``naked" AGN as discovered by Hawkins (2004) are among
the best candidates to look for ``true" unabsorbed Seyfert 2 galaxies (Panessa \& Bassani 2002).
Indeed, the monitored optical variability over 25 years points
to a nucleus seen directly where, given the absence of broad optical emission lines 
in the optical spectra, the BLR may lack. 
As successfully proven for NGC\,3147 (Bianchi et al. 2008), 
the key strategy to spot this class of AGN is through simultaneous optical and
X-ray observations in order to rule out variability as the cause
of the optical and X-ray mismatch. 
We have therefore observed two ``naked" AGN candidates, 
Q\,2130-431 and Q\,2131-427, simultaneously in the X-ray and optical band.

\subsection{Q\,2131-427: a ``true" unabsorbed Seyfert 2 galaxy}

For Q\,2131-427, the photometric measurements obtained over $\sim$ 25 years, 
as reported in Hawkins (2004), show a slow decrease of the emission, from B $\sim$ 20 mag 
in 1974 to B $\sim$ 21 mag in 2002. The Optical Monitor on board XMM-\textit{Newton} (Mason et al. 2001) 
measured B = 20.83 $\pm$ 0.10, only slightly above the last measurement in the B$_J$ light curve (in 2001)
B$_J$ = 21.2 $\pm$ 0.3, confirming that the source is still in a low flux state. 
Interestingly, Q\,2131-427 shows no signs of broad emission line
components in its optical spectrum and the faint X-ray spectrum of this source is unabsorbed, with an upper limit on the intrinsic column density of
$N_{\rm H, int}$ $<$ 9 $\times$ $10^{20}$ cm$^{-2}$, consistent with that from a host galaxy. This picture is hard to reconcile within the Unified Model predictions, where the BLR is obscured by a
toroidal absorbing medium of dust and gas, the former responsible of the optical extinction of the broad emission lines and the latter measurable in
X-rays. 

One possibility is that the BLR is hidden behind a Compton thick medium 
whose column density cannot be measured in X-ray spectra below 10 keV (Matt et al. 1999). 
The upper limit on the EW of the FeK$_{\alpha}$ line at 6.4 keV does not rule out the Compton thick hypothesis, however this measurement is not
reliable, given the low statistics spectrum above 5 keV. Instead, the F$_{2-10 keV}$/F$_{[\textsc{oiii}]}$ 
ratio\footnote{F$_{[\textsc{oiii}]}$ has been reddening corrected (Bassani et al. 1999).} of $\sim$ 6 is typical of Compton thin
sources. 

Tran (2001) studied the properties of type 2 Seyfert galaxies with polarized (hidden) broad-line regions (HBLR) seen in reflected light 
compared to the non-HBLR Seyfert 2 galaxies, introducing the use of a diagnostic diagram involving [O\textsc{iii}]$\lambda$5007/H$_{\beta}$. 
The median values of [O\textsc{iii}]$\lambda$5007/H$_{\beta}$ found are 6.8 $\pm$ 1.5 and 9.9 $\pm$ 1.3 for 
the non-HBLR and HBLR Seyfert 2 galaxies respectively. The [O\textsc{iii}]$\lambda$5007/H$_{\beta}$ ratio of Q\,2131-427 is $\sim$ 9, suggesting that this
source might harbor a hidden BLR. In this hypothesis, some sort of dusty medium should obscure the optical broad emission lines 
and be gas-free.

As already invoked for the unabsorbed Seyfert 2 in Panessa \& Bassani (2002), a dust-to-gas ratio 10-50 times higher than Galactic would
be enough to obscure the BLR. However, this is an unlikely scenario as a dust obscuration is often observed to be lower than the associated 
gas column density (see Maiolino et al. 2001). Moreover, the Balmer decrement of the NLR, i.e. (H$_{\alpha}$/H$_{\beta}$)$_{NLR}$ $\sim$ 4.2, 
suggests that the environment is not particularly dusty (Ward et al. 1987)
and the optical extinction A$_{V}$ $\sim$ 1.3 mag derived from the Balmer decrement is still too small to 
be able to completely obscure a standard BLR. 

The hypothesis of a pc and/or sub-pc scale clumpy obscuring medium which results in a variable column density
in X-ray spectra (Risaliti et al. 2002, Elvis et al. 2004, Risaliti et al. 2005, Bianchi et al. 2009) 
cannot help to explain the complete absence
of broad components in the optical spectrum. Under the assumption that the X-ray absorption originates from 
the BLR clouds (see Risaliti et al. 2009), we should expect variable column density in the X-ray spectra
but still broad emission lines in the optical spectrum, indeed at these spatial scales the gas should be dust-free
being within the sublimation radius.

The most likely explanation for the observed optical/X-ray discrepancy is to assume that in this source
the BLR does not exist or its emission is gradually declining. Indeed, the well documented optical flux decrease may 
favor the hypothesis that the BLR fade away in response to a decrease of the continuum emission. 
Recent models propose a dependence of the BLR properties with luminosity and/or accretion rate (Wang \& Zhang 2007, Elitzur \& Shlosman 2006, Nicastro et al. 2003,
Nicastro 2000, Williams et al. 1999).
According to Elitzur \& Shlosman (2006), the classical molecular torus is just the outer part of a clumpy wind that 
comes out off the accretion disk. At bolometric luminosities lower than $\sim$ $10^{42}$ ergs sec$^{-1}$, the mass accretion
can no longer sustain the cloud outflow rate causing the disappearance of the torus and, at lower luminosities, of the broad line clouds too.
According to the above scheme, the absence of a BLR in Q\,2131-427 is not justified 
given the high bolometric luminosity (L$_{Bol}$ $\sim$ 8 $\times$ 10$^{44}$ ergs sec$^{-1}$); however
the low Eddington ratio measured (L$_\mathrm{Bol}$/L$_\mathrm{Edd}$ = 0.0024) may favor the scenario
in which the BLR is formed in a vertical disk wind originating at a critical distance in the accretion disk,
at accretion rates higher than a minimum value ($\sim 10^{-3}$, Nicastro 2000). According to this model,
below the critical accretion rate the BLR cannot be formed. In the case of Q\,2131-427, 
this is one of the best hypothesis to interpret the observational mismatch, taking into account that
the Eddington ratio is around this critical value and it is probably affected by uncertainties related to the determination
of the black hole mass and of the bolometric correction. 
If the BLR properties actually depend on the luminosity and/or accretion physics, this must have strong 
implications for AGN Unification Models and also for population studies of obscured and unobscured AGN samples (e.g. fraction of type 2
objects versus luminosity) as discussed in Elitzur \& Shlosman (2006) 
(see also Wang \& Zhang (2007) for an evolutionary sequence of Seyfert galaxies).

\subsection{Q\,2130-431: a ``true" intermediate Seyfert galaxy?}

The Q\,2130-431 optical spectrum shown here reveals that this source is 
a type 1.8 Seyfert galaxy, i.e. we detect H$_{\alpha}$ and
H$_{\beta}$ broad line components (see Figure~\ref{opt1}). The H$_{\alpha}$ broad component was not detected by Hawkins (2004) since, 
at the redshift of the source, the line is shifted out of their optical spectra. Also
the H$_{\beta}$ broad component was not detected, maybe because it is
difficult to identify given its width, being hidden under a galaxy continuum (see Figure 6 in Hawkins 2004).
Alternatively, it may be variable. Note that the FWHM of the H$_{\beta}$ narrow line component has not 
varied significantly between 2002 ($\sim$ 426 km s$^{-1}$) and 2006 (320$^{+50}_{-30}$ km s$^{-1}$) observations. 

The simultaneous ultraviolet imaging with the Optical Monitor, allows us to detect the source in the UVW1 filter (245-320 nm), with
a Galactic reddening corrected magnitude of $20.11 \pm 0.07$, slightly brighter when compared to the 
last measurement of $B_J=20.35 \pm 0.05$ mag, as shown in the 25 years light curve (see Fig. 8, Hawkins 2004).
In X-rays, the 2-10 keV luminosity of $\sim$ 4 $\times$ 10$^{43}$ ergs s$^{-1}$ suggests that
this is a typical type 1 Seyfert galaxy with a standard photon index ($\Gamma$ = 1.8-2; e.g., Piconcelli et al. 2005) and
no intrinsic absorption ($<$ 2 $\times$ $10^{20}$ cm$^{-2}$, at 90\% confidence level). 
The F$_{2-10 keV}$/F$_{[\textsc{oiii}]}$ ratio is $\sim$ 77, well within the regime of Compton thin objects
(Cappi et al. 2006, Bassani et al. 1999). Our X-ray flux is in agreement with the \textit{Chandra} one found by 
Gliozzi et al. (2007) suggesting no particular variability in X-rays.

According to the above observational evidence, Q\,2130-431 should not be considered as a pure
``naked" AGN, in the sense that it displays optical broad emission line components and, 
therefore, it has a BLR. However, intermediate type Seyfert galaxies are characterized
by large Balmer decrements consistent with reddening of the BLR and determining
the 1.8/1.9 spectroscopic type (Osterbrock 1981). The weakness of the broad Balmer lines is often
ascribed to dust absorption (Maiolino \& Rieke 1995) and, indeed, moderate absorption is observed in X-rays (Risaliti et al. 1999).
In Q\,2130-431, the optical BLR Balmer decrement of 3.4 (almost the same value of the NLR Balmer decrement) suggests 
that the BLR does not suffer from heavy reddening. In addition, in the X-ray spectrum, no absorption has been measured,
strengthening the idea that the BLR is not obscured and that the weakness of the broad optical lines can be intrinsic,
e.g. due to a small amount of gas in the clouds. A weak BLR can be due to a luminosity/accretion rate related disappearance, 
as suggested for Q\,2131-427. 
Interestingly the existence of this kind of objects should be hypothesized within an evolutionary sequence, as 
the one drawn in Elitzur \& Shlosman (2006), from type 1 objects to ``true" type 2s.
However, the high X-ray luminosity and the Eddington ratio (L$_{Bol}$/L$_{Edd}$ $\sim$ 0.37) disfavor the latter hypothesis. 
Alternatively, the lack of a prominent BLR in Q\,2130-431 may simply be a peculiar intrinsic property of this object.
In both scenarios, Q\,2130-431 can represent a case of ``true" intermediate Seyfert galaxy, in which the BLR is intrinsically
weak instead of being obscured by dust or gas.

\section{Conclusions}

We have obtained quasi-simultaneous X-ray (XMM-Newton) and optical (NTT-EMMI)
spectra of two Seyfert galaxies, Q\,2130-431 and Q\,2131-427, belonging to the class of ``naked" AGN characterized
by Hawkins (2004) and, currently, composed of six members. The strong optical brightness variability and the lack
of broad emission line components in their spectra have suggested that
in these AGN the nucleus is seen directly and it is taken off its BLR.
Our quasi-simultaneous observations have proven fundamental to confirm
the ``naked" hypothesis in the case of Q\,2131-427, for which we have ruled out
the Compton thick nature based on the low F$_{2-10 keV}$/F$_{[\textsc{oiii}]}$ ratio. Moreover
the Balmer decrement in the NLR excludes a high dust-to-gas ratio as a possible explanation
of the broad emission lines absence. Instead, the present observations have confirmed
that no broad components are found associated to the optical spectrum emission lines
and simultaneously no absorption is measured in the X-ray spectrum. Either a dusty medium is absorbing
the BLR being adequately gas free to let the X-ray continuum escape, however this is an unlikely scenario, or the source
is accreting below a critical accretion rate such that the BLR cannot be formed (Nicastro 2000).

The ``naked" hypothesis is instead discarded for Q\,2130-431, which indeed 
displays H$_{\alpha}$ and H$_{\beta}$ broad line components in its optical spectrum. However, 
no sign of mild absorption in its X-ray continuum is found, contrary to what expected for intermediate Seyfert galaxies 
(in order to justify the partial obscuration of the BLR), neither the BLR itself seems to suffer from heavy reddening. 
The presence of a small amount of gas to photoionize can be the reason for
the intermediate optical spectrum, making it a case of a potential ``true" intermediate Seyfert galaxy.

At last, in the appendix, we present the results on the cluster Abell 3783 serendipitously detected in the Q\,2131-427
EPIC FOV providing a description of its fundamental parameters.

\section*{Acknowledgments}
We thank the anonymous referee for the useful comments.
F.P. acknowledges support by ASI-INAF I/08/07/0 grants. FJC and XB acknowledge
support by the Spanish Ministry of Science and Innovation, under grants ESP2006-13608-C02-01.
This work benefited from an Italian-Spanish Integrated Action (HI 2006-0079).
F.P. thanks M.T. Ceballos for the technical support.
P.J. Humphrey is thanked for the use of his surface brightness fitting code. 
This research has made use of data obtained from the
High Energy Astrophysics Science Archive Research Center (HEASARC),
provided by NASA's Goddard Space Flight Center. This research has
also made use of the NASA/IPAC Extragalactic Database (\ned) which is
operated by the Jet Propulsion Laboratory, California Institute of
Technology, under contract with NASA.

\appendix

\section[]{The serendipitous detection of X-ray emission by the cluster Abell 3783}

\begin{figure*}
\includegraphics[width=0.5\textwidth,angle=-90]{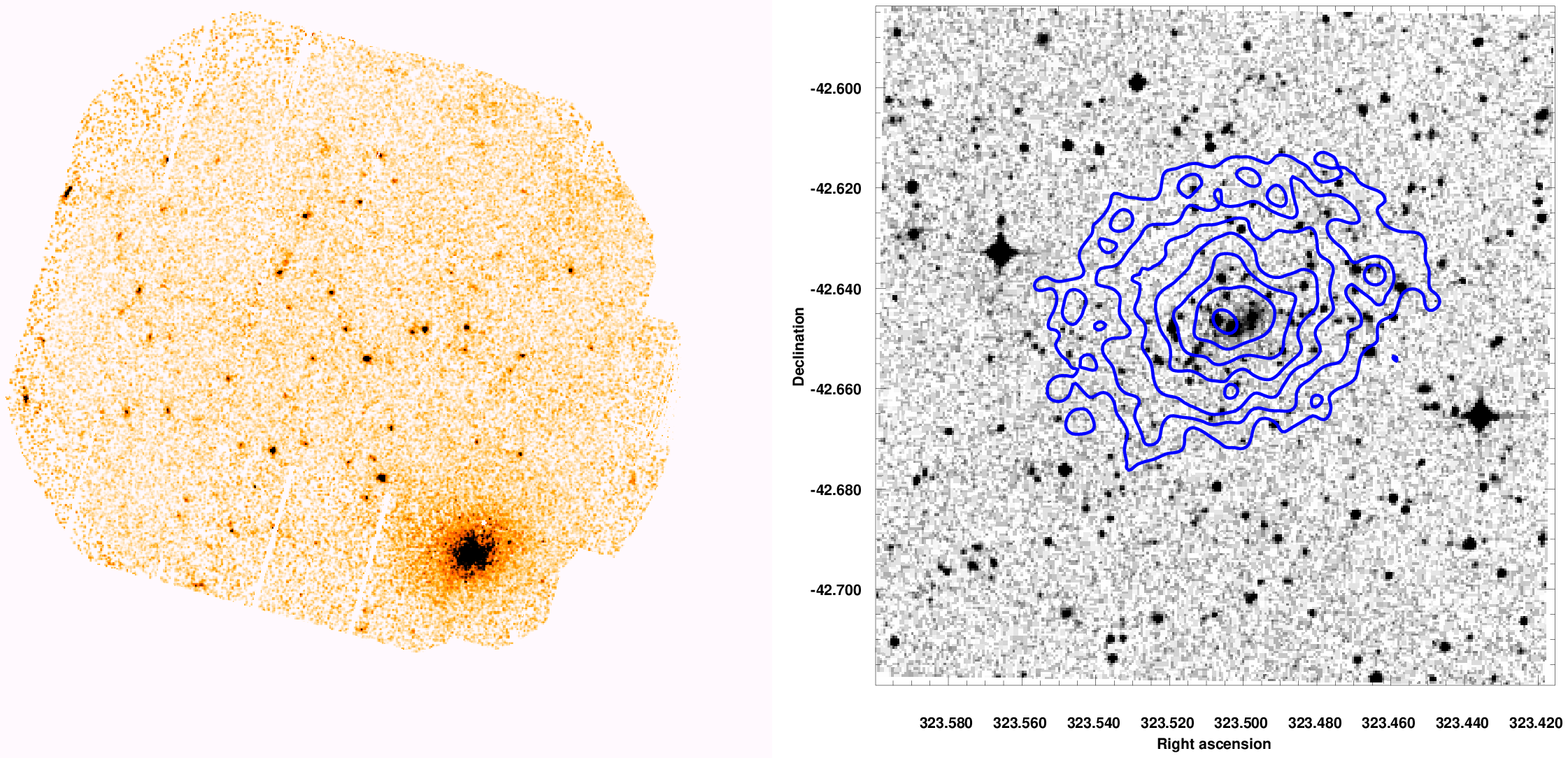}
\caption{\emph{Left:}Exposure corrected 0.5-2.0 keV combined MOS1, MOS 2 and pn X-ray
image of the field of Q\,2131-427.
The cluster is clearly visible as the extended source near the edge of the field of view.
\emph{Right:}The DSS image of the cluster with X-ray contours overlaid and logarithmically
spaced between $5 \times 10^{-6}$ counts/s/pixel and $5 \times 10^{-5}$ counts/s/pixel. Equatorial
coordinates are given at J2000 equinox.
}
\label{fig.1}
\end{figure*}

An extended source, of a $\sim3$ arc-minute scale, in the field of Q\,2131-427 
is clearly detected at 10.2\arcmin\ off-axis (see Fig.\ref{fig.1}). 
The X-ray peak of
of the source in equatorial coordinates is
\mbox{$\alpha_{\rm J2000.0}=21^{\rm h}34^{\rm m}00^{\rm s},\, \delta_{\rm J2000.0}=-42^{\circ}38^{\rm m}46^{\rm s}$} and
it is associated with the cluster of galaxies \source\ (see the overlay of the X-ray contours on the DSS image in Fig.\ref{fig.1}) at the redshift of 
$z=0.195$ (West \& Fradsen 1981). 
With the caveat of the \xmm\ PSF at this off-axis angle 
($\sim 8$\arcsec\ is the 50\% encircled energy fraction radius at 1.5 keV) the
peak is consistent with being associated with the brightest galaxy (BCG) 
of the cluster, 2MASX J21335944-4238448, rather than the likely interloper
2MASX J21340071-4238528 with position 
\mbox{$\alpha_{\rm J2000.0}=21^{\rm h}34^{\rm m}00.7^{\rm s},\, \delta_{\rm J2000.0}=-42^{\circ}38^{\rm m}53^{\rm s}$} at a redshift $z=0.028$. These
are the only galaxies with redshifts available listed in \ned\ within 
3\arcmin\ of the source.
Considerations of the redshift derived from the Fe K$\alpha$ complex in the 
X-ray spectrum, the luminosity and the temperature of the cluster give
further strength to the association. In the following we will therefore
assume $z=0.195$ as the redshift of the source for which 1\arcmin\ corresponds to 194 kpc. 
All the errors quoted in this Appendix are at the 68\% confidence limit.

The data reduction and preparation follow the procedure of 
Gastaldello et al. (2008) where more details can be found; here
we provide just a brief description. The data were re-processed with SAS v8.0.

%
\begin{figure*}
\begin{center} \includegraphics[width=0.4\textwidth,angle=-90]{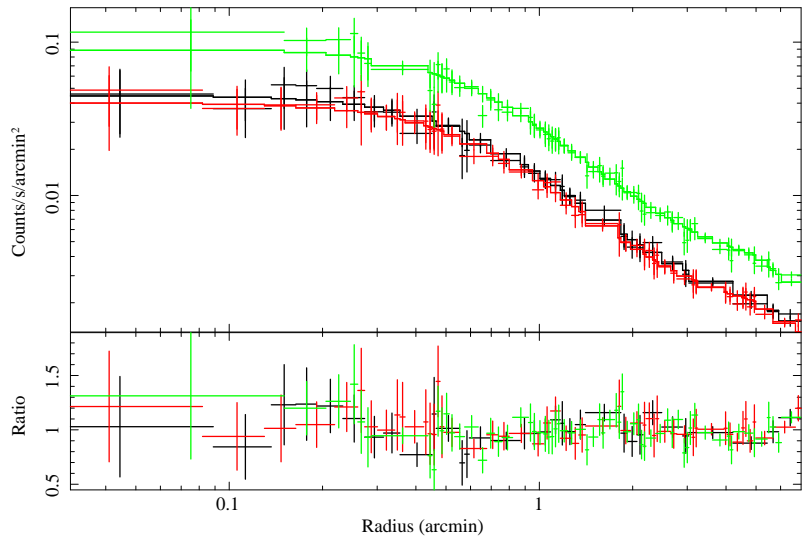}
\caption{Surface brightness profile of the X-ray emission of \source. 
Data from MOS1, MOS2 and pn are plotted in black, red and green respectively. 
The best fit beta model and ratio of data over the model are also shown.}
\label{fig.2} \end{center}
\end{figure*}
%
\begin{figure*}
  \begin{center} \includegraphics[width=0.4\textwidth,
  angle=-90]{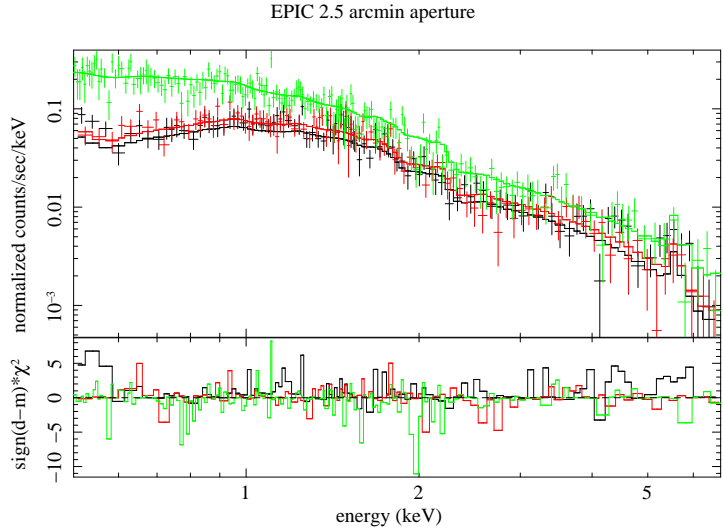} \caption{X-ray spectrum of the \source\ 
taken from a 2.5\arcmin\ aperture centered on the peak of the emission. Data 
from MOS1, MOS2 and pn are plotted in black, red and green respectively. The 
best fit model and residuals are also shown.}
  \label{fig.3} \end{center}
\end{figure*}
%
\begin{figure*}
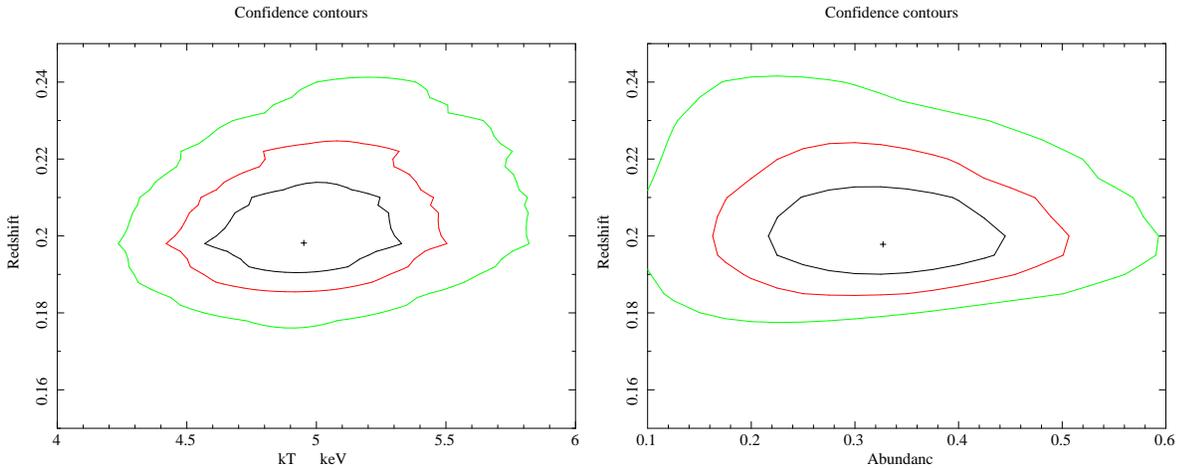

\begin{center} 
\includegraphics[width=0.35\textwidth,angle=-90]{contour_z_T.ps} 
\includegraphics[width=0.35\textwidth,angle=-90]{contour_z_Z_angr.ps}
\caption{The 68\%, 90\% and 99\% redshift-temperature $\chi^{2}$ contours on the 
left panel and redshift-abundance on the right panel derived from the spectrum of
Fig.\ref{fig.3}}
  \label{fig.4} \end{center}
\end{figure*}

For each detector we created images in the 0.5-2 keV band with point sources 
masked using circular regions of 25\arcsec\ radius centered at the source 
position. The images have been exposure corrected and a radial surface 
brightness profile was extracted from a circular region of 
7\arcmin\ of radius centered on the cluster centroid. 
We account for the X-ray background in the surface brightness analysis 
by including a constant-background component. The data were grouped to
have at least 20 counts per bin in order to apply the $\chi^{2}$ statistic.
The fitted model is convolved with the \xmm\ PSF. The joint best-fit 
$\beta$-model (Cavaliere \& Fusco Femiano 1976) has a core radius of $r_c = 120\pm13$ 
kpc ($37$\arcsec$\pm4$\arcsec) and $\beta=0.54\pm0.03$ for a 
$\chi^{2}$/d.o.f. = 210/141 (see Fig.\ref{fig.2}). Fits to the profiles of 
the individual detectors give consistent results within 1$\sigma$ of the 
combined-fit result and in the case of the MOS2 detector is formally 
acceptable (49/47). The main contribution to the
$\chi^{2}$ comes mainly from the pn ($\chi^{2}$/d.o.f. = 102/61 for the fit 
to the individual profile) and its 
origin is instrumental (CCD gaps
and pixels under-sampling the PSF), whereas a large area of the radial profile 
at $r > 5$\arcmin\ is lost for MOS1 ($\chi^{2}$/d.o.f. = 55/29) due to the missing CCD-6,
 hence there is no need for more complicated models.

For spectral fitting, we extracted spectra for each detector from a 2.5\arcmin\
region centered on the peak of the emission, to maximize the S/N over the
background. Redistribution matrix files (RMFs) and
ancillary response files (ARFs) were generated using the SAS tasks
{\em rmfgen} and {\em arfgen} in extended source
mode with appropriate flux weighting. The background was estimated
locally using spectra extracted from a source free region of the same extent 
at the same off-axis angle (avoiding CCD-4 for MOS1 and CCD-5 for MOS 2 
because in this observation they display anomalous background for energies below 1 keV, 
see Kuntz \& Snowden 2008). 
The spectra from the three 
detectors were re-binned to ensure a signal-to-noise
ratio of at least 3 and a minimum 20 counts per bin and they were
jointly fitted with an \apec\ thermal plasma modified by Galactic
absorption (Dickey \& Lockman 1990). The spectral fitting was performed 
in the 0.5-7 keV band. The spectra are shown in 
Fig.\ref{fig.3}: 
the best fit parameters are $kT = 4.94\pm0.25$ keV and  
$Z=0.32\pm0.02$ \solar\ for a $\chi^{2}$/d.o.f. = 418/396.
The source photons correspond to about 78\% of the total events ($\sim 3400$
counts in each MOS and 5700 in the pn).
If we leave the redshift parameter free we obtain $z=0.20\pm0.01$, in good agreement with the 
redshift determined optically: the quality of the \xmm\ data is such to tightly constrain the 
redshift of the cluster; in Fig.\ref{fig.4} we show the temperature-redshift and 
redshift-metal abundance $\chi^{2}$ confidence contours.

A likely previous X-ray identification is the \rosat\ source 1RXS J213401-423833
in the \rosat\ All-Sky bright source catalogue (Voges et al. 1999) and associated
with \source. The \rosat\ count-rate of $6.32\pm1.61\times10^{-2}$ ct/s which
corresponds to an unabsorbed flux in the 0.1-2.4 keV band of $1.04\pm0.30\times10^{-12}$ 
\fxunits\ (for a 5\arcmin\ extraction radius) is in good agreement with our
determination of $1.30\pm0.05\times10^{-12}$ in the 0.1-2.4 keV band (according 
to the derived surface brightness model the 2.5\arcmin-5\arcmin\ region contributes
only $\sim 25$\% of the flux within a 5\arcmin\ region).

To investigate possible spatial variation in
the spectral parameters of the cluster, we extracted two annular
regions of radii 0\arcmin-1\arcmin\ and 1\arcmin-3\arcmin. The
derived spectral parameters are: $kT = 4.78\pm0.28$ keV and $Z=0.51\pm0.13$ 
\solar\ with $\chi^{2}$/d.o.f. = 226/177 for the inner annulus; 
$kT=4.50^{+0.40}_{-0.35}$ keV and $Z=0.09^{+0.11}_{-0.09}$ \solar\ with 
$\chi^{2}$/d.o.f. = 313/314 for the outer annulus. 
The width of the bins have been chosen in order to avoid bias in the 
temperature measurement caused by scattered flux by the PSF. 
It is difficult, given the off-axis position of the cluster, to asses the
existence of gradients in the temperature: given the quality of the data
the cluster is consistent with being isothermal over the explored radial range.
The abundance gradient, signature of relaxed cool-core cluster (e.g., De Grandi \& Molendi 2001), 
is on the contrary significant at the 2.6$\sigma$ level.

Under the assumption of isothermality and assuming that the cluster follows the
best fit model to the surface brightness profile derived above  we can calculate the
total mass profile using the best-fit $\beta$-model for which
the gas density and total mass profiles can be expressed by simple
analytical formula (e.g., Ettori 2000).
We evaluated $r_{500}$ as the radius at which the density is 500 times the
the critical density and the virial radius as the radius at which the
density corresponds to $\Delta_{\rm{vir}}$, as obtained by Bryan \& Norman (1998)
\footnote{$\Delta_{vir}=18\pi^2+82x-39x^2$ where $x=\Omega(z)-1$, $\Omega(z)=\Omega_m(1+z)^3/E(z)^2$ and $E(z)=\left[\Omega_m(1+z)^3+\Lambda\right]^{1/2}$} 
for the concordance cosmological model used in this paper.
To evaluate the errors on the estimated quantities we used the same procedure
as above repeating the measurements for  10000 random selections drawn from
Gaussian distributions for the temperature and parameters of the surface 
brightness profile.
For $\Delta=500$ we obtained,
$M_{500} = (2.69\pm0.31) \times 10^{14}$ \msun\ within
$r_{500} = 920\pm36$ kpc;
the virial mass is, $M_{\rm{vir}}= (5.65\pm0.65)\times10^{14}$ \msun,
within the virial radius $r_{\rm{vir}} = 1911\pm73$ kpc.
If the identification of 2MASX J21335944-4238448 with the BCG is correct, the optical
luminosity calculated in the $K_S$ following Lin \& Mohr 2004 is $M_{K_S}=-27.02$
corresponding to $1.40\times10^{12}$ \lsun, consistent with the central galaxy luminosity-host
halo mass relation (and its scatter) derived in the literature (e.g.,  Lin \& Mohr 2004).
We calculated the gas mass to be $M_{gas,500} = (3.37\pm0.37) \times 10^{13}$ \msun.
For \source, the aperture of
485 kpc used for spectroscopy encloses 74\% of the flux within $r_{500}$. 
The derived bolometric
luminosity within $r_{500}$ is $L_{500} = (3.61\pm0.26) \times 10^{44}$\lxunits: \source\
lies very close to the \lx-\tx\ relation derived from the cluster sample of
objects with \tx$>3$ keV of Markevitch (1998), though given the angular resolution
of the data it has not been possible to correct for the effect of an eventual presence of a cool 
core (Allen \& Fabian 1998; Markevitch 1998).

Finally, as done routinely in our previous analysis (e.g., Gastaldello et al. 2007, 2008)
we have studied the sensitivity of our results to various systematic uncertainties and data analysis 
choices that may impact our results, like varying \nh\ or using a different energy band
(1-7 keV) or standard blank background fields for background subtraction, finding always agreement
within the errors with the results reported above. Below we summarize two possible sources
of errors not previously explored in our work.

{\it Cross-calibration:} improvements have been made in the status of the
cross-calibration between the EPIC cameras in the recent SAS releases, but
some calibration differences are still present which can affect the spectral
parameters of an extended source positioned off-axis like \source\ in the
present observation. In the  2.5\arcmin\ aperture the fit to the individual MOS1 
data returns $kT = 5.42^{+0.54}_{-0.41}$, $Z=0.59^{+0.20}_{-0.18}$ and \xspec\ 
$norm = 1.98\pm0.11\times10^{-3}$ ($\chi^{2}$/d.o.f. = 111/105); for MOS 2 
$kT = 4.44^{+0.44}_{-0.39}$, $Z=0.24^{+0.15}_{-0.14}$ and  
$norm = 1.93\pm0.10\times10^{-3}$ ($\chi^{2}$/d.o.f. = 96/114); for pn
 $kT = 4.54^{+0.40}_{-0.36}$, $Z=0.25^{+0.12}_{-0.11}$ and  
$norm = 1.77\pm0.07\times10^{-3}$($\chi^{2}$/d.o.f. = 176/186). 
The above results correspond to an unabsorbed flux in the 0.5-2 keV band
of $8.91\pm0.70\times10^{-13}$ \fxunits\ and $L_{500} = (4.38\pm0.47) \times 10^{44}$ 
\lxunits\ for MOS1,
$8.09\pm0.60\times10^{-13}$ \fxunits\ and  $L_{500} = (3.49\pm0.36) \times 10^{44}$ \lxunits\
for MOS2, $7.40\pm0.48\times10^{-13}$ \fxunits\ and $L_{500} = (3.22\pm0.31) \times 10^{44}$ 
\lxunits\ 
for pn. The results are within 1$\sigma$ of the best joint fit and the overall trend of 
higher fluxes of the MOS compared to the pn is consistent with the current calibration
uncertainties (Guainazzi 2008, Mateos et al. 2009).

{\it Radial range used in the surface brightness fitting:} Simulations and 
simple analytic models pointed out how the $\beta$ model
overestimates gas mass because it returns a biased low $\beta$ due to the
restricted range of radii where the fit is performed
(e.g. Bartelmann \& Steinmetz 1996; Roncarelli et al. 2006). Indeed recent analyses 
investigating surface brightness profiles of clusters with \rosat\ and 
\chandra\ find evidence for a steepening of the gas density slope with radius 
for clusters (Vikhlinin et al. 1999, 2006; Ettori \& Balestra 2009) .
We investigated fitting the surface brightness profile in a narrower range 
(5\arcmin) and wider range (9\arcmin) compared to the choice made in the
analysis. Using the narrower range we obtain a core radius of $r_c = 90\pm16$ 
kpc ($28$\arcsec$\pm5$\arcsec) and $\beta=0.46\pm0.03$ for a 
$\chi^{2}$/d.o.f. = 149/129 and using the wider range we obtain 
$r_c = 162\pm16$ kpc ($50$\arcsec$\pm5$\arcsec) and $\beta=0.65\pm0.04$ for a 
$\chi^{2}$/d.o.f. = 338/156. The use of a radial range as large as possible 
is clearly preferred in the determination of the $\beta$ model parameters,
but we did not adopt the fit obtained in the 9\arcmin\ aperture because of the 
large residuals due to a not optimal fitting of the background level.
This is due to the fact that, because of the off-axis position of the source, 
the increased radial range is obtained with only a partial azimuthal coverage 
due to the increasing out of field of view area (only $\sim40$\% of the 6\arcmin-9\arcmin annulus is covered by data in the MOS1 and pn detectors). Had we used the parameters
obtained in the narrower range we would have derived  
$M_{500} = (2.13\pm0.27) \times 10^{14}$ \msun\ within $r_{500} = 851\pm36$ kpc
and $M_{\rm{vir}}= (4.45\pm0.56)\times10^{14}$ \msun,
within the virial radius $r_{\rm{vir}} = 1765\pm73$ kpc whereas using the
wider range we would have derived  
$M_{500} = (3.50\pm0.43) \times 10^{14}$ \msun\ within $r_{500} = 1005\pm41$ kpc
and $M_{\rm{vir}}= (7.44\pm0.90)\times10^{14}$ \msun,
within the virial radius $r_{\rm{vir}} = 2095\pm84$ kpc.

\end{document}